\documentclass[runningheads]{llncs}

\usepackage{graphicx}
\usepackage{amsmath}
\usepackage{epsfig} 
\usepackage{mathptmx} 
\usepackage{times} 
\usepackage{amssymb}  
\usepackage{amsmath,bm}
\usepackage{etoolbox}
\makeatother
\usepackage{multirow}
\usepackage[caption=false, font=footnotesize]{subfig}
\usepackage{xcolor}
\usepackage{hyperref}
\begin{document}
\title{Unsupervised Anomaly Detection in Medical Images Using Masked Diffusion Model}
%
\author{Hasan Iqbal\inst{1*}\orcidID{0009-0005-2162-3367} \and
Umar Khalid\inst{2*}\orcidID{0000-0002-3357-9720}\thanks{* Equal Contribution} \and
Chen Chen\inst{2}\orcidID{0000-0003-3957-7061} \and
Jing Hua\inst{1}\orcidID{0000-0002-3981-2933}
}
\authorrunning{F. Author et al.}

\institute{Department of Computer Science, Wayne State University, Detroit, MI, USA \and
Center for Research in Computer Vision, University of Central Florida, Orlando, FL, USA
}
\maketitle              
\begin{abstract}
It can be challenging to identify brain MRI anomalies using supervised deep-learning techniques due to anatomical heterogeneity and the requirement for pixel-level labeling. Unsupervised anomaly detection approaches provide an alternative solution by relying only on sample-level labels of healthy brains to generate a desired representation to identify abnormalities at the pixel level. Although, generative models are crucial for generating such anatomically consistent representations of healthy brains, accurately generating the intricate anatomy of the human brain remains a challenge. In this study, we present a method called the masked-denoising diffusion probabilistic model (mDDPM), which introduces masking-based regularization to reframe the generation task of diffusion models. Specifically, we introduce Masked Image Modeling (MIM) and  Masked Frequency Modeling (MFM) in our self-supervised approach that enables models to learn visual representations from unlabeled data. To the best of our knowledge, this is the first attempt to apply MFM in denoising diffusion probabilistic models (DDPMs) for medical applications. We evaluate our approach on datasets containing tumors and numerous sclerosis lesions and exhibit the superior performance of our unsupervised method as compared to the existing fully/weakly supervised baselines. Project website: \url{https://mddpm.github.io/}.

\end{abstract}
\section{Introduction}
 \begin{figure*}
    \includegraphics[trim={0 0cm 0 0},clip,scale = 0.27]{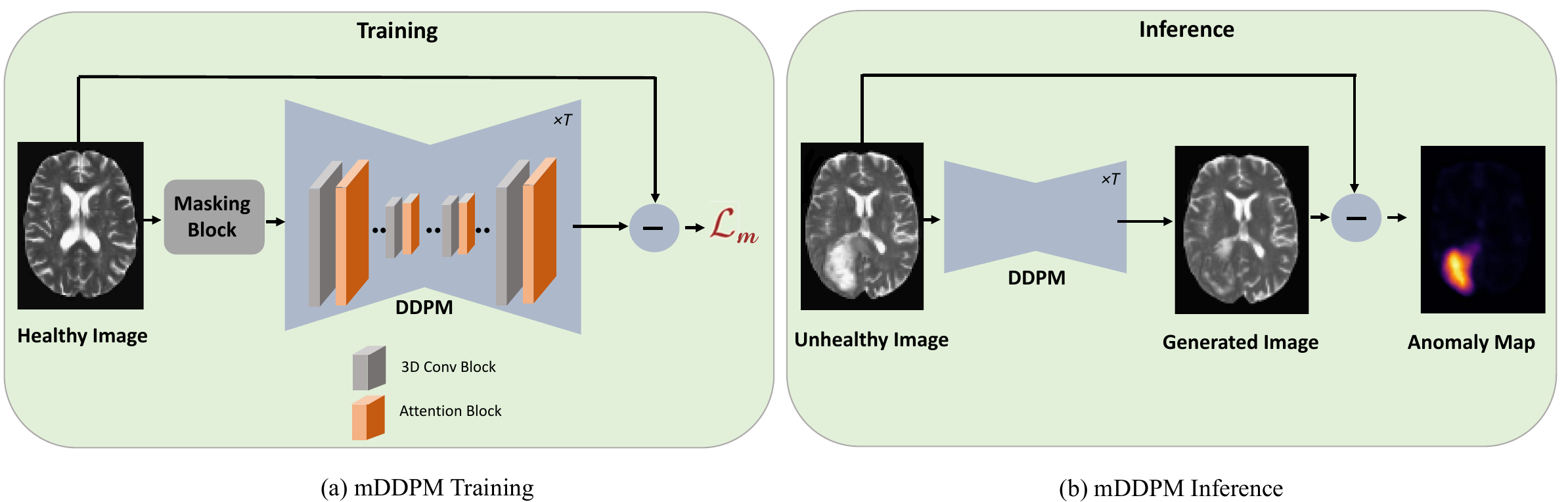}
\centering
\caption{\footnotesize Schematic diagram of our framework. \textbf{(a)} During training, only healthy images are used, and no classifier guidance is required. The healthy image is passed through the \textbf{Masking Block} before feeding into the DDPM. The reconstruction loss is calculated between the original image and the image generated by the DDPM. Here, \textbf{Masking Block} plays the role of regularizer and eliminates the need for classifier guidance which is discussed in Section~\ref{sec:method}. Further, the schematic of the \textbf{Masking Block} is illustrated in Fig.~\ref{fig2}. \textbf{(b)} During inference, DDPM considers the tumor in the unhealthy image as an augmented patch and eliminates it to generate a healthy image. The difference between the generated image and the given unhealthy image is then calculated to report the anomaly map. No masking mechanism is employed at inference.}
\label{fig:method}
\end{figure*}
Medical imaging (MI) systems play a crucial role in aiding radiologists in their diagnostic and decision-making processes. These systems provide medical imaging specialists with detailed visual information to detect abnormalities, make accurate diagnoses, plan treatments, and monitor patients. More recently, advanced machine learning techniques and image processing algorithms are being utilized to automate the medical diagnostic process. Among these techniques, deep learning models based on convolutional neural networks (CNNs) have exhibited significant achievements in accurately identifying anomalies in medical images~\cite{shen2017deep,10.1145/3464423}. However, supervised CNN approaches have inherent limitations, including the requirement for extensive expert-annotated training data and the difficulty of learning from noisy or imbalanced data \cite{ELLIS2022100068,karimi2020deep,Johnson.2019}. On the contrary, pixel-level annotations are not necessary for unsupervised anomaly detection (UAD) that uses only healthy examples for training. 

In recent years, numerous architectures have been explored to investigate UAD for brain MRI anomaly detection. Autoencoders (AE) and variational autoencoders (VAE) have proven to be effective in training models and achieving efficient inference. However, their reconstructions often suffer from blurriness, limiting their effectiveness in UAD~\cite{baur2021autoencoders}. To address this limitation, researchers have focused on enhancing the understanding of image context by utilizing the spatial context through techniques such as spatial erasing~\cite{zimmerer2018context} and leveraging 3D information~\cite{bengs2021three,behrendt2022capturing}. Additionally, vector-quantized VAEs~\cite{pinaya2022unsupervised}, adversarial autoencoders~\cite{Chen.2018}, and encoder activation maps~\cite{SILVARODRIGUEZ2022102526} have been proposed to improve the image restoration quality. Generative adversarial networks (GANs) have emerged as an alternative to AE-based architectures for the task of UAD~\cite{Schlegl.2019}. However, the unstable training nature of GANs poses challenges, and GANs often suffer from mode collapse and a lack of anatomical coherence~\cite{baur2021autoencoders,nguyen2021unsupervised}.

Recently, denoising diffusion probabilistic models (DDPMs)~\cite{ho2020denoising} have shown promise for UAD in brain MRI~\cite{wyatt2022anoddpm,behrendt2023patched,pinaya2022fast}. In the context of DDPMs, the approach involves introducing noise to an input image and subsequently utilizing a trained model to eliminate the noise and estimate or reconstruct the original image~\cite{ho2020denoising}. Unlike most autoencoder-based methods, DDPMs retain spatial information in their hidden representations, which is crucial for the image generation process~\cite{rombach2022high}. Recent works in medical imaging establish that they exhibit scalable and stable training properties and generate high-quality, sharp images with classifier guidance~\cite{wolleb2022diffusion,wyatt2022anoddpm,sanchez2022healthy,pinaya2022fast,pinaya2022unsupervised}. Further,~\cite{behrendt2023patched,ozdenizci2023restoring,lugmayr2022repaint} introduce patch-based DDPM (pDDPM) which offer better brain MRI reconstruction by incorporating global context information about individual brain structures and appearances while estimating individual patches.

Given the advantages observed when applying Mask Image Modeling (MIM) in conjunction with VAE frameworks \cite{wang2023fremae,gao2022convmae,he2021masked}, such as enhanced generalization capabilities and the acquisition of a comprehensive understanding of the structural characteristics of images, we introduce the first investigation into leveraging  Masked Image Modeling (MIM) and Masked Frequency Modeling (MFM) within DDPMs. In our proposed framework, masked DDPM (mDDPM), the need for the classifier guidance is eliminated. The masking mechanism proposed in our framework serves as a unique regularizer that enables the incorporation of global information while preserving fine-grained local features. This regularization technique imposes a constraint on DDPM, ensuring the generation of a healthy image during inference, regardless of the input image characteristics. In this study, we focus on exploring three specific variants of masking-based regularization: \textit{(i) image patch-masking (IPM)}, \textit{(ii) frequency patch-masking (FPM)}, and \textit{(iii) frequency patch-masking CutMix (FPM-CM)}. 
In the IPM approach, random pixel-level masks are applied to patches extracted from the original image before subjecting them to the diffusion process in DDPMs. The unmasked version of the same image is used as the reference for comparison. However, in the FPM approach, the image is first transformed into the frequency domain using the Fast Fourier Transform. Subsequently, patch-level masking is performed in the frequency domain as shown in Figure~\ref{2a}. The inverse Fourier Transform is then applied to obtain the reconstructed image, which is utilized to calculate the reconstruction loss. In FPM-CM, random patches are sampled from the augmented image generated through FPM and subsequently inserted at corresponding positions within the original clean image as shown in Figure~\ref{2b}.
To evaluate the performance of our method,  we use two publicly available datasets: BraTS21~\cite{baid2021rsna}, and MSLUB~\cite{lesjak2018novel}, and demonstrate a significant improvement (p $<$ 0.05) in tumor segmentation performance.
\vspace{-2mm}
\section{Method} \label{sec:method}
\vspace{-2mm}
 Given the potential occurrence of anomalies in any region of the brain during testing, we introduce data augmentation techniques that involve the insertion of random augmented patches into the healthy input image, $\bm{z} \in \mathbb{R}^{C,W,H}$ with $C$ channels, width $W$ and height $H$,  prior to the application of DDPM noise addition and removal. This approach allows us to generate a healthy image based on an unhealthy image during inference, facilitating the calculation of the anomaly map. The illustration of our approach can be seen in Figure \ref{fig:method}. We will be further discussing our unsupervised mDDPM approach and proposed masking strategies in this section. 
\subsection{Fourier Transform}
We first briefly introduce the Discrete Fourier Transform (DFT) as it plays a crucial role in our mDDPM approach. 
Given a 2D signal $\bm{z} \in \mathbb{R}^{W \times H}$, the corresponding 2D-DFT, a widely used signal analysis technique, can be defined as follows:

\begin{equation}
	\small
	f(x, y)=\sum_{h=0}^{H-1} \sum_{w=0}^{W-1} {z}{(h, w)} e^{-j 2 \pi\left(\frac{xh}{H}+\frac{yw}{W}\right)},\label{eq:fft}
\end{equation}

 $z{(h, w)}$ denotes the signal located at position $(h, w)$ in $\bm{z}$, while $x$ and $y$ serve as indices representing the horizontal and vertical spatial frequencies in the Fourier spectrum.
The inverse 2D DFT (2D-IDFT) is defined as:
\begin{equation}
	\small
	F(h, w)=\frac{1}{H W} \sum_{x=0}^{H-1} \sum_{y=0}^{W-1} f(x, y) e^{j 2 \pi\left(\frac{xh}{H}+\frac{yw}{W}\right)},\label{eq:ifft}
\end{equation}

Both the DFT and IDFT can be efficiently computed using the Fast Fourier Transform (FFT) algorithm, as in~\cite{nussbaumer1981fast}.

In the context of medical images with various modalities, the Fourier Transform is applied independently to each channel. Additionally, previous works such as~\cite{wang2023fremae,chen2019drop,kauffmann2014neural} have demonstrated that the high-frequency part of the Fourier spectrum contains detailed structural texture information, while the low-frequency part contains global information.

\vspace{-3mm}

\subsection{DDPMs}In DDPMs, the forward process involves gradually introducing noise to the input image $\bm{z}_0$ according to a predefined schedule $\beta_1,...,\beta_T$. The noise is sampled from a Gaussian distribution $\mathcal{N}(\textbf{0},\textbf{I})$, and at each time step $t$, the noisy image $\bm{z}_t$ is generated as follows:

\begin{equation}
\bm{z}_t \sim q(\bm{z}_t|\bm{z}_0)=\mathcal{N}(\sqrt{\bar\alpha_t} \bm{z}_0,(1-\bar\alpha_t) \textbf{I}),
\end{equation}

where $\bar\alpha_t=\prod\nolimits_{s=0}^{t}(1-\beta_t)$ and $t$ is sampled from a uniform distribution. For $t=T$, the image becomes pure Gaussian noise $\bm{z}_t = \bm{\epsilon} \sim \mathcal{N}(\textbf{0},\textbf{I})$, while for $t=0$, $\bm{z}_t$ remains $\bm{x}_0$.

In the denoising process, the objective is to reverse the forward process and reconstruct the original image $\bm{z}_0$. The reconstruction is achieved by sampling $\bm{z}_0$ from the joint distribution:

\begin{equation}
\bm{z}_{0} \sim p_{\theta}(\bm{z_t})\prod\nolimits^T_{t=1}p_{\theta}(\bm{z}_{t-1}|\bm{z}_t),
\end{equation}

where $p_{\theta}(\bm{z}_{t-1}|\bm{z}_t)$ is modeled as a Gaussian distribution $\mathcal{N}(\bm{\mu}_{\theta}(\bm{z}_t,t),\bm{\Sigma}_{\theta}(\bm{z}_t,t))$. The parameters $\bm{\mu}_{\theta}$ and $\bm{\Sigma}_{\theta}$ are estimated by a neural network, and we use a U-Net architecture for this purpose. The covariance $\bm{\Sigma}_{\theta}(\bm{z}_t,t)$ is fixed as $\frac{1-\alpha_{t-1}}{1-\alpha_t} \beta_t \textbf{I}$, following the approach in \cite{ho2020denoising}.
In this work, we simplify the loss derivation by directly estimating the reconstruction $\bm{z}^{'}_0 \sim p_{\theta}(\bm{z}_{0}|\bm{z}_t,t)$ as in ~\cite{behrendt2023patched} and using the mean absolute error ($l1$-error) as the loss function:

\vspace{-3mm}

\begin{equation} 
\mathcal{L}_{rec} = |\bm{z}_0 - \bm{z}^{'}_{0}|,\label{eq:lec}
\end{equation}

Instead of performing step-wise denoising for all time steps starting from $t=T$, as commonly done for sampling images with DDPMs, we directly estimate $\bm{z}^{'}_0$ at a fixed time step $t_{fix}$.

\vspace{-3mm}

\subsection{Masked DDPMs}

As stated above, we model mDDPM by introducing a Masking block before DDPM stage as shown in Fig.~\ref{fig:method} that can be incorporated in three different forms during training, namely IPM, FPM, and FPM-CM. 
\subsubsection{Image Patch-Masking (IPM)} 
\begin{figure}[b!] 
    \centering
  \subfloat[ Block diagram of FPM. Here, Fast Fourier Transform is performed on each slice, and then masking is performed on the frequency spectrum. The Inverse Fourier Transform of this masked spectrum outputs an augmented image of the input image.\label{2a}]{%
       \includegraphics[trim={0 1cm 0 0},width=0.49\textwidth]{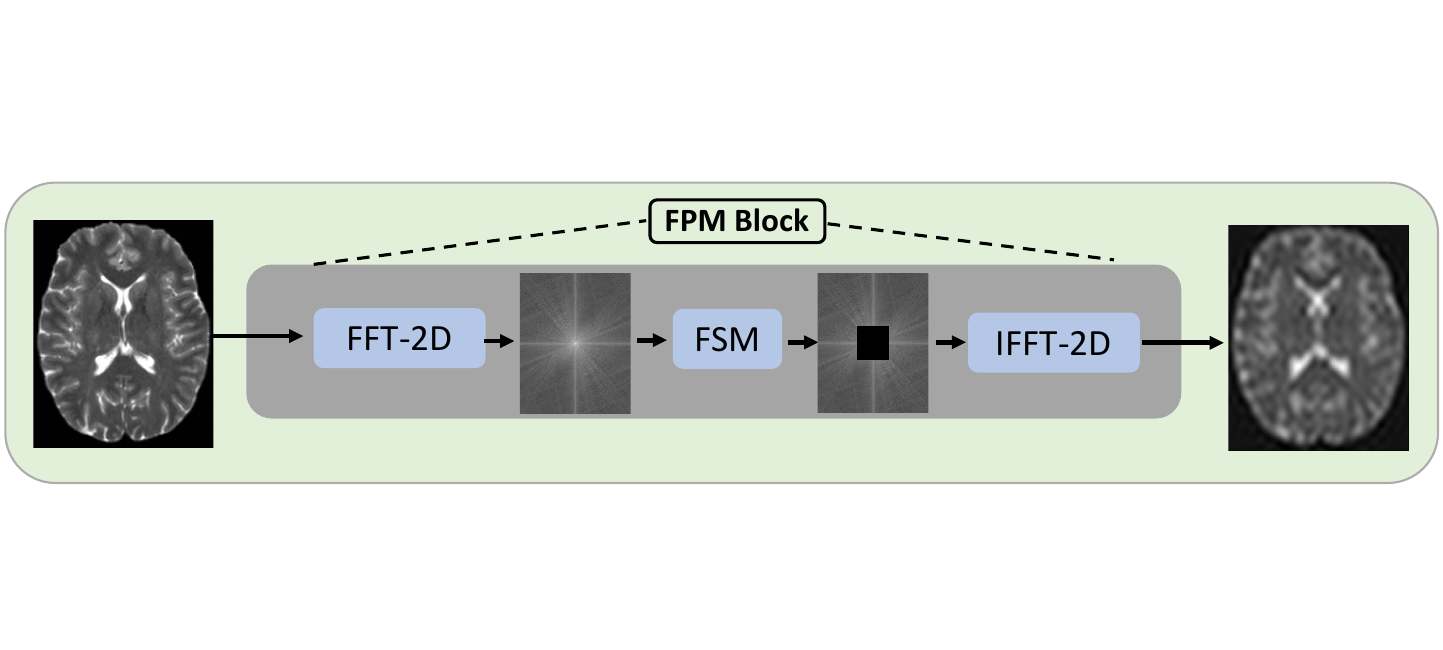}}
     \hfill
  \subfloat[ Block diagram of FPM-CM. Here, random patches are chosen from the augmented image obtained after FPM. These selected patches are then inserted at the corresponding positions in the original clean image to generated an augmented image. \label{2b}]{%
        \includegraphics[width=0.49\textwidth]{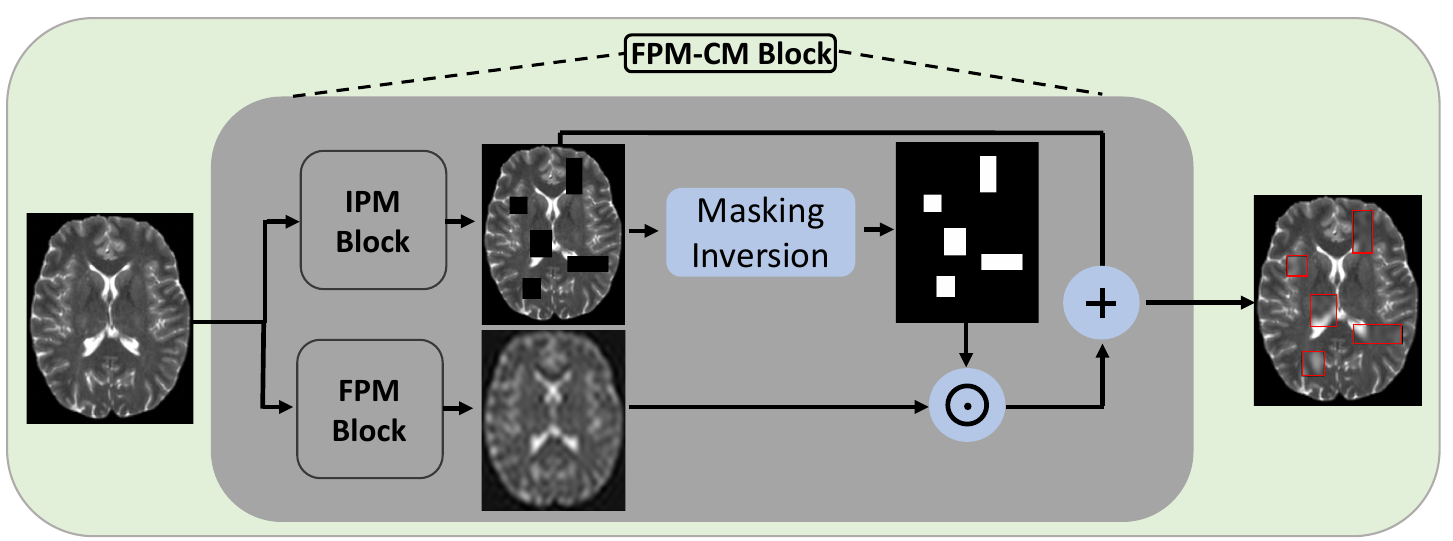}}
    \hfill
  \caption{Proposed data augmentation techniques which are implemented in the Masking Block of the Fig.~\ref{fig:method}.}
  \label{fig2} 
\end{figure}

In the IPM approach, random masks are applied at the pixel level to patches extracted from the original image. The masked image is then subjected to the diffusion process in DDPMs. In contrast, the unmasked version of the same image is used as a reference for comparison during training. For reference, the output of the IPM block can be observed in Fig.~\ref{2b}.

During training, we sample $N$ patch regions, $[p_1,p_2,...,p_N]$ at random positions such that $\Sigma_{n=0}^{N} {A_{n}} < A_{z}$, where ${A_{n}}$ is the area of patch $p_n$, and ${A_{z}}$ is the area of image, $\bm{z}$ in pixel space.
Let $\bm{M^I_p} \in \mathbb{R}^{C,H,W}$ be a binary mask in the pixel domain that indicates which pixels overlap with the patches $[p_1,p_2,...,p_N]$. Specifically, the pixels within each $\bm{p_n}$ are assigned a value of zero, while the pixels outside of $\bm{p_n}$ are assigned a value of one. We obtain $\bm{z_M}$, by combining the original image $\bm{z_0}$ with the masked region using element-wise multiplication:
\begin{equation}
\bm{z^I_M} =\bm{z}_0 \odot \bm{M^I_p}, 
\end{equation}
Here, $\odot$ represents the Hadamard product. $\bm{z_M}$ is then fed to DDPM forward process.

In the backward process, the denoised image, denoted as $\bm{\tilde{z}}_0$, is generated by the network as an estimate of the original image $\bm{z}_0$. As we calculate the absolute difference between the original image $\bm{z}_0$ and the denoised image $\bm{\tilde{z}_0}$ during training, the objective function $\mathcal{L}_{M} =\mathcal{L}_{rec}$, where $\mathcal{L}_{rec}$ is defined in Eq.~\ref{eq:lec}. 

By applying random pixel-level masks to the patches, the IPM approach introduces a form of regularization that encourages the DDPMs to generate images that closely resemble the unmasked reference image. 
\vspace{-4mm}
\subsubsection{Frequency Patch-Masking (FPM)}The proposed FPM mechanism has been illustrated in Fig.~\ref{2a}. FPM block mainly consists of three blocks, FFT-2D, frequency spectrum masking (FSM), and IFFT-2D. Here, the image undergoes a series of transformations in the frequency domain using the Fast Fourier Transform (FFT).
The Fast Fourier Transform allows us to analyze the image in terms of different frequencies present within it. After the image is transformed into the frequency domain, patch-level masking is performed using a binary mask $\bm{M^F_p}$ similar to the one used in IPM. This means that specific regions or patches within the frequency representation of the image are masked out. Masking in the frequency domain allows for selective modification of certain frequency components of the image while leaving others intact. Here, we don't specify the high-frequency or low-frequency masking, rather MFM is performed randomly. Once the patch-level masking is complete, the inverse Fourier Transform is applied to the modified frequency representation. This transforms the image back from the frequency domain to the spatial domain, reconstructing the modified image. The FPM block is mathematically formulated as,
\begin{equation}
    \bm{z^F_M}=IDFT(\bm{M^F_p} \odot DFT(\bm{z_0})), 
\end{equation}
The reconstructed image is subsequently inputted into DDPM, utilizing the identical objective function as described earlier in Eq.~\ref{eq:lec}.
\subsubsection{Frequency Patch-Masking CutMix (FPM-CM)}FPM-CM extends the application of FPM to apply patched augmentation in the original image as shown in Fig.~\ref{2b}. In FPM-CutMix, the image is first passed through an FPM augmentation stage, and then patches are sampled from this augmented image. These patches are essentially small rectangular regions with varying sizes that capture specific features or information from the augmented image.
After sampling the patches, they are inserted into the original clean image at corresponding positions. This means that the patches are placed in the same spatial locations within the original image as they were in the augmented image. By inserting the patches at corresponding positions, the intention is to transfer the modified features from the augmented image to the original clean image. It can be observed that these augmented patches behave as anomalies during training as the key idea is to use DDPM to generate the clean un-augmented image. Therefore such augmentation serves the purpose of unsupervised anomaly training. Assuming that we have obtained $\bm{z^I_M}$, and $\bm{z^F_M}$ from IPM and FPM block respectively, FPM-CM can be mathematically written as,
 \begin{equation}
    \bm{z^{FC}_M}= \bm{z^I_M} + (\bm{z^F_M} \odot \neg \bm {M^I_p}), 
\end{equation}
Here, $\neg$ indicates binary inversion, and $\bm{z^{FC}_M}$ is the FPM-CM block output. We feed in DDPM with $\bm{z^{FC}_M}$ using the same objective function as used by other approaches mentioned above and stated in Eq.~\ref{eq:lec}.
\section{Experimental Evaluation}
\subsection{Implementation Details}
For our experiments, we utilize the publicly available IXI dataset for training \cite{ixii}. The IXI dataset comprises 560 pairs of T1 and T2-weighted brain MRI scans. To ensure robust evaluation, the IXI dataset is partitioned into eight folds, comprising 400/160 training/validation samples. To evaluate our approach, we employ two publicly available datasets: the Multimodal Brain Tumor Segmentation Challenge 2021 (BraTS21) dataset \cite{baid2021rsna} and the multiple sclerosis dataset from the University Hospital of Ljubljana (MSLUB) \cite{lesjak2018novel}. The BraTS21 dataset includes 1251 brain MRI scans with four different weightings (T1, T1-CE, T2, FLAIR). Following~\cite{behrendt2023patched}, it was divided into a validation set of 100 samples and a test set of 1151 samples, both containing unhealthy scans. The MSLUB dataset consists of brain MRI scans from 30 multiple sclerosis (MS) patients, with each patient having T1, T2, and FLAIR-weighted scans. This dataset was split into a validation set of 10 samples and a test set of 20 samples as in ~\cite{behrendt2023patched}, all representing unhealthy scans. Thus, our training, validation, and testing data consist of 400, 270, and 1171  samples respectively. In all our experimental setups, we exclusively employ T2-weighted images extracted from the respective dataset and perform the pre-processing such as  affine transformation, skull stripping, and downsampling as in ~\cite{behrendt2023patched}. With the specifically designed pre-processing techniques, we filtered out the regions belonging to the foreground area so that Masking Block can only be applied to the foreground pixel patches. 

We assess the performance of our proposed method, mDDPM, in comparison to various established baselines for UAD in brain MRI. These baselines include: \textit{(i)} VAE \cite{baur2021autoencoders}, \textit{(ii)}  Sequential VAE (SVAE) \cite{behrendt2022capturing}, \textit{(iii)} denoising AE (DAE) \cite{kascenas2021denoising}, the GAN-based \textit{(iv)} f-AnoGAN \cite{Schlegl.2019}, \textit{(v)} DDPM \cite{wyatt2022anoddpm}, and \textit{(iv)} patched DDPM (pDDPM), which feeds patched input to the DDPM. We evaluate all baseline methods via in-house training using their default parameters. For VAE, SVAE, and f-AnoGAN, we set the value of the hyperparameter according to~\cite{behrendt2023patched}. For DDPM, pDDPM, and mDDPM, we employ structured simplex noise instead of Gaussian noise as it better captures the natural frequency distribution of MRI images \cite{wyatt2022anoddpm}. We follow~\cite{behrendt2023patched} for all other training and inference settings. By default, the models undergo training for 1600 epochs.  During the training phase, the volumes are processed in a slice-wise fashion, where slices are uniformly sampled with replacement. However, during the testing phase, all slices are iterated over to reconstruct the entire volume. Further, we conducted all our experiments with a masking ratio randomly varying between 10\%-90\% of the whole foreground region.
\subsection{Inference Criteria} During the training phase, all models are trained to minimize the $l1$ error between the healthy input image and its corresponding reconstruction. At the test stage, we utilize the reconstruction error as a pixel-wise anomaly score denoted as $\bm{\Lambda}_{S}=|\bm{z}_0-\bm{z}^{'}_0|$, where higher values correspond to larger reconstruction errors. 

To enhance the quality of the anomaly maps, we employ commonly used post-processing techniques \cite{baur2021autoencoders,zimmerer2018context}. Prior to binarizing $\bm{\Lambda}_{S}$, we apply a median filter with a kernel size of $K_M=5$ to smooth the anomaly scores and perform brain mask erosion for three iterations. After binarization and calculating threshold as in ~\cite{behrendt2023patched}, we iteratively calculate DICE scores~\cite{dice1945measures} for different thresholds and select the threshold that yields the highest average DICE score on the selected test set. Additionally, we record the average Area Under the Precision-Recall Curve (AUPRC)~\cite{davis2006relationship} on the test set.
 \begin{table*}[t]
\caption{\footnotesize{Comparison of the models under consideration, with the best outcomes denoted in bold using DICE and AUPRC as evaluation metrics.}}
 \centering
\begin{tabular}{l|c|c|c|c}
\hline
\multirow{2}{*}{\textbf{Model}}& \multicolumn{2}{c}{BraTS21} & \multicolumn{2}{c}{MSLUB}  \\
 & \textbf{DICE [\%]} & \textbf{AUPRC [\%]} & \textbf{DICE [\%]} & \textbf{AUPRC [\%]} \\
\hline
\textit{VAE} \cite{baur2021autoencoders} & 30.57$ \pm $1.67 & 28.47$ \pm $1.38 & 6.63$ \pm $0.12 & 5.01$ \pm $0.54 \\
\textit{SVAE} \cite{behrendt2022capturing} & 33.86$ \pm $0.19 & 33.53$ \pm $0.23 & 5.71$ \pm $0.48 & 5.05$ \pm $0.11\\
\textit{DAE} \cite{kascenas2021denoising} & 36.85$ \pm $1.62 & 45.19$ \pm $1.35 & 3.67$ \pm $0.82 & 5.24$ \pm $0.53  \\
\textit{f-AnoGAN} \cite{Schlegl.2019} & 24.44$ \pm $2.28 & 22.52$ \pm $2.37 & 4.29$ \pm $1.02 & 4.09$ \pm $0.79  \\
\textit{DDPM} \cite{wyatt2022anoddpm} & 40.82$ \pm $1.34 & 49.82$ \pm $1.13 & 8.52$ \pm $1.42 & 8.44$ \pm $1.54  \\

\textit{pDDPM~\cite{behrendt2023patched}} + \textit{fixed sampling} + $\mathcal{L}_{p}$ & 49.12$ \pm $1.27 & 53.98$ \pm $2.16 & 9.04$ \pm $0.66 & 9.23$ \pm $0.83  \\
\hline
\textit{mDDPM (IPM)} &  {52.16$ \pm $1.64} & {58.12$ \pm $1.56}&  {10.39$ \pm $0.88} & {10.58$ \pm $0.92} \\
\textit{mDDPM (FPM)} &  {50.91$ \pm $1.28} & {56.27$ \pm $1.44}&  9.31$ \pm $0.46 & 9.51$ \pm $0.52 \\
\textit{mDDPM (FPM-CM)} &  \textbf{53.02$ \pm $1.34} & \textbf{59.04$ \pm $1.26}&  \textbf{10.71$ \pm $0.62} & \textbf{10.59$ \pm $0.57}  \\
\hline
\end{tabular}

\label{tab:mainresults}
\end{table*}
 \begin{figure}[t!]
    \includegraphics[trim={0 0cm 0 0},clip,scale = 0.20]{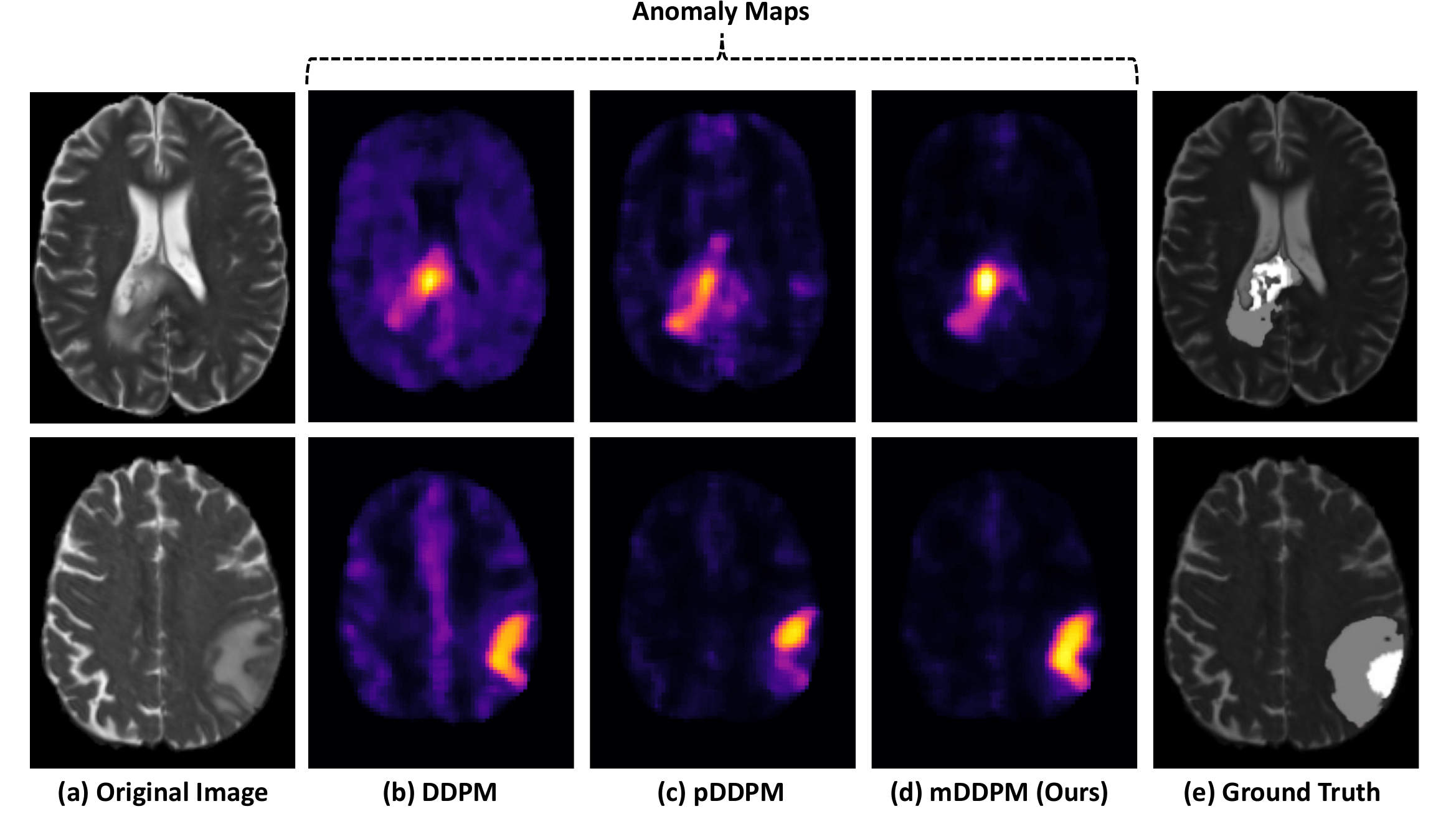}
\centering
\caption{\footnotesize In the comparison of the qualitative results between mDDPM (FPM-CM) and the baselines, we present the anomaly map comparisons for two samples. In the first sample, mDDPM demonstrates a more precise detection of the anomaly, exhibiting a final anomaly map without foreground noise. Similarly, in the second sample, mDDPM performs closest to the ground truth in terms of anomaly detection as shown in the results of the second row where we only included results for the FPM-CM variant of our method.}
\label{fig:results}
\vspace{-1em}
\end{figure}
\subsection{Results}
The comparison of our proposed mDDPM with the baseline method is presented in Table~\ref{tab:mainresults}. It can be observed that our mDDPM outperforms all baseline approaches on both datasets in terms of DICE~\cite{dice1945measures} and AUPRC~\cite{davis2006relationship}. In terms of qualitative evaluation, we observe smaller reconstruction errors from mDDPM compared to patched DDPM~\cite{behrendt2023patched} for healthy brain anatomy, as shown in Fig.~\ref{fig:results}. It can be observed that, mDDPM (FPM-CM) showcases a higher level of precision in detecting the anomaly, resulting in an anomaly map that is free from foreground noise.
\vspace{-2mm}
\section{Conclusion}
\vspace{-2mm}
This study introduces an approach for reconstructing the healthy brain anatomy using masked DDPM, which incorporates image-mask and frequency-mask regularization. Our method, known as mDDPM, surpasses established baselines, even with unsupervised training. However, a limitation of the proposed approach is the increased inference time associated with the diffusion architecture. To address this, future research could concentrate on enhancing the efficiency of the diffusion denoising process by leveraging spatial context more effectively. Further, we intend to explore the Masked Diffusion Transformer architecture in our future studies where we can incorporate a latent modeling scheme using masks to specifically improve the contextual relationship learning capabilities of DDPMs for object semantic parts within an image.

%
%
%
%





\bibliographystyle{splncs04}

\bibliography{template/references}

\begin{thebibliography}{10}
\providecommand{\url}[1]{\texttt{#1}}
\providecommand{\urlprefix}{URL }
\providecommand{\doi}[1]{https://doi.org/#1}

\bibitem{ixii}
\url{https://brain-development.org/ixi-dataset/}

\bibitem{baid2021rsna}
Baid, U., Ghodasara, S., Mohan, S., Bilello, M., Calabrese, E., Colak, E.,
  Farahani, K., Kalpathy-Cramer, J., Kitamura, F.C., Pati, S., et~al.: The
  rsna-asnr-miccai brats 2021 benchmark on brain tumor segmentation and
  radiogenomic classification. arXiv preprint arXiv:2107.02314  (2021)

\bibitem{baur2021autoencoders}
Baur, C., Denner, S., Wiestler, B., Navab, N., Albarqouni, S.: Autoencoders for
  unsupervised anomaly segmentation in brain mr images: a comparative study.
  Medical Image Analysis p. 101952 (2021)

\bibitem{behrendt2022capturing}
Behrendt, F., Bengs, M., Bhattacharya, D., Kr{\"u}ger, J., Opfer, R.,
  Schlaefer, A.: Capturing inter-slice dependencies of 3d brain mri-scans for
  unsupervised anomaly detection. In: Medical Imaging with Deep Learning (2022)

\bibitem{behrendt2023patched}
Behrendt, F., Bhattacharya, D., Kr{\"u}ger, J., Opfer, R., Schlaefer, A.:
  Patched diffusion models for unsupervised anomaly detection in brain mri.
  arXiv preprint arXiv:2303.03758  (2023)

\bibitem{bengs2021three}
Bengs, M., Behrendt, F., Kr{\"u}ger, J., Opfer, R., Schlaefer, A.:
  Three-dimensional deep learning with spatial erasing for unsupervised anomaly
  segmentation in brain mri. International journal of computer assisted
  radiology and surgery  \textbf{16}(9),  1413--1423 (2021)

\bibitem{Chen.2018}
Chen, X., Konukoglu, E.: {Unsupervised Detection of Lesions in Brain MRI using
  Constrained Adversarial Auto-encoders}. In: {International Conference on
  Medical Imaging with Deep Learning (MIDL)}. {Proceedings of Machine Learning
  Research}, PMLR (2018)

\bibitem{chen2019drop}
Chen, Y., Fan, H., Xu, B., Yan, Z., Kalantidis, Y., Rohrbach, M., Yan, S.,
  Feng, J.: Drop an octave: Reducing spatial redundancy in convolutional neural
  networks with octave convolution. In: Proceedings of the IEEE/CVF
  international conference on computer vision. pp. 3435--3444 (2019)

\bibitem{davis2006relationship}
Davis, J., Goadrich, M.: The relationship between precision-recall and roc
  curves. In: Proceedings of the 23rd international conference on Machine
  learning. pp. 233--240 (2006)

\bibitem{dice1945measures}
Dice, L.R.: Measures of the amount of ecologic association between species.
  Ecology  \textbf{26}(3),  297--302 (1945)

\bibitem{ELLIS2022100068}
Ellis, R.J., Sander, R.M., Limon, A.: Twelve key challenges in medical machine
  learning and solutions. Intelligence-Based Medicine  \textbf{6},  100068
  (2022)

\bibitem{10.1145/3464423}
Fernando, T., Gammulle, H., Denman, S., Sridharan, S., Fookes, C.: Deep
  learning for medical anomaly detection – a survey. ACM Comput. Surv.
  \textbf{54}(7) (jul 2021). \doi{10.1145/3464423},
  \url{https://doi.org/10.1145/3464423}

\bibitem{gao2022convmae}
Gao, P., Ma, T., Li, H., Lin, Z., Dai, J., Qiao, Y.: Convmae: Masked
  convolution meets masked autoencoders (2022)

\bibitem{he2021masked}
He, K., Chen, X., Xie, S., Li, Y., Dollár, P., Girshick, R.: Masked
  autoencoders are scalable vision learners (2021)

\bibitem{ho2020denoising}
Ho, J., Jain, A., Abbeel, P.: Denoising diffusion probabilistic models.
  Advances in Neural Information Processing Systems  \textbf{33},  6840--6851
  (2020)

\bibitem{Johnson.2019}
Johnson, J.M., Khoshgoftaar, T.M.: {Survey on deep learning with class
  imbalance}. {Journal of Big Data}  \textbf{6}(1),  1--54 (2019)

\bibitem{karimi2020deep}
Karimi, D., Dou, H., Warfield, S.K., Gholipour, A.: Deep learning with noisy
  labels: Exploring techniques and remedies in medical image analysis. Medical
  Image Analysis  \textbf{65},  101759 (2020)

\bibitem{kascenas2021denoising}
Kascenas, A., Pugeault, N., O'Neil, A.Q.: Denoising autoencoders for
  unsupervised anomaly detection in brain mri. In: {International Conference on
  Medical Imaging with Deep Learning (MIDL)}. {Proceedings of Machine Learning
  Research}, PMLR (2022)

\bibitem{kauffmann2014neural}
Kauffmann, L., Ramano{\"e}l, S., Peyrin, C.: The neural bases of spatial
  frequency processing during scene perception. Frontiers in integrative
  neuroscience  \textbf{8}, ~37 (2014)

\bibitem{lesjak2018novel}
Lesjak, {\v{Z}}., Galimzianova, A., Koren, A., Lukin, M., Pernu{\v{s}}, F.,
  Likar, B., {\v{S}}piclin, {\v{Z}}.: A novel public mr image dataset of
  multiple sclerosis patients with lesion segmentations based on multi-rater
  consensus. Neuroinformatics  \textbf{16}(1),  51--63 (2018)

\bibitem{lugmayr2022repaint}
Lugmayr, A., Danelljan, M., Romero, A., Yu, F., Timofte, R., Van~Gool, L.:
  Repaint: Inpainting using denoising diffusion probabilistic models. In:
  Proceedings of the IEEE/CVF Conference on Computer Vision and Pattern
  Recognition. pp. 11461--11471 (2022)

\bibitem{nguyen2021unsupervised}
Nguyen, B., Feldman, A., Bethapudi, S., Jennings, A., Willcocks, C.G.:
  Unsupervised region-based anomaly detection in brain mri with adversarial
  image inpainting. In: 2021 IEEE 18th international symposium on biomedical
  imaging (ISBI). pp. 1127--1131. IEEE (2021)

\bibitem{nussbaumer1981fast}
Nussbaumer, H.J., Nussbaumer, H.J.: The fast Fourier transform. Springer (1981)

\bibitem{ozdenizci2023restoring}
{\"O}zdenizci, O., Legenstein, R.: Restoring vision in adverse weather
  conditions with patch-based denoising diffusion models. IEEE Transactions on
  Pattern Analysis and Machine Intelligence  (2023)

\bibitem{pinaya2022fast}
Pinaya, W.H., Graham, M.S., Gray, R., Da~Costa, P.F., Tudosiu, P.D., Wright,
  P., Mah, Y.H., MacKinnon, A.D., Teo, J.T., Jager, R., et~al.: Fast
  unsupervised brain anomaly detection and segmentation with diffusion models.
  arXiv preprint arXiv:2206.03461  (2022)

\bibitem{pinaya2022unsupervised}
Pinaya, W.H., Tudosiu, P.D., Gray, R., Rees, G., Nachev, P., Ourselin, S.,
  Cardoso, M.J.: Unsupervised brain imaging 3d anomaly detection and
  segmentation with transformers. Medical Image Analysis  \textbf{79},  102475
  (2022)

\bibitem{rombach2022high}
Rombach, R., Blattmann, A., Lorenz, D., Esser, P., Ommer, B.: High-resolution
  image synthesis with latent diffusion models. In: Proceedings of the IEEE/CVF
  Conference on Computer Vision and Pattern Recognition. pp. 10684--10695
  (2022)

\bibitem{sanchez2022healthy}
Sanchez, P., Kascenas, A., Liu, X., O’Neil, A.Q., Tsaftaris, S.A.: What is
  healthy? generative counterfactual diffusion for lesion localization. In:
  MICCAI Workshop on Deep Generative Models. pp. 34--44. Springer (2022)

\bibitem{Schlegl.2019}
Schlegl, T., Seeb{\"o}ck, P., Waldstein, S.M., Langs, G., Schmidt-Erfurth, U.:
  {f-AnoGAN: Fast unsupervised anomaly detection with generative adversarial
  networks}. {Medical image analysis}  \textbf{54},  30--44 (2019)

\bibitem{shen2017deep}
Shen, D., Wu, G., Suk, H.I.: Deep learning in medical image analysis. Annual
  review of biomedical engineering  \textbf{19}, ~221 (2017)

\bibitem{SILVARODRIGUEZ2022102526}
Silva-Rodríguez, J., Naranjo, V., Dolz, J.: Constrained unsupervised anomaly
  segmentation. Medical Image Analysis  \textbf{80},  102526 (2022)

\bibitem{wang2023fremae}
Wang, W., Wang, J., Chen, C., Jiao, J., Sun, L., Cai, Y., Song, S., Li, J.:
  Fremae: Fourier transform meets masked autoencoders for medical image
  segmentation (2023)

\bibitem{wolleb2022diffusion}
Wolleb, J., Bieder, F., Sandk{\"u}hler, R., Cattin, P.C.: Diffusion models for
  medical anomaly detection. arXiv preprint arXiv:2203.04306  (2022)

\bibitem{wyatt2022anoddpm}
Wyatt, J., Leach, A., Schmon, S.M., Willcocks, C.G.: Anoddpm: Anomaly detection
  with denoising diffusion probabilistic models using simplex noise. In:
  Proceedings of the IEEE/CVF Conference on Computer Vision and Pattern
  Recognition. pp. 650--656 (2022)

\bibitem{zimmerer2018context}
Zimmerer, D., Kohl, S., Petersen, J., Isensee, F., Maier-Hein, K.:
  Context-encoding variational autoencoder for unsupervised anomaly detection.
  In: International Conference on Medical Imaging with Deep Learning--Extended
  Abstract Track (2019)

\end{thebibliography}


@article{vernooij2007incidental,
  title={Incidental findings on brain MRI in the general population},
  author={Vernooij, Meike W and Ikram, M Arfan and Tanghe, Herv{\'e} L and Vincent, Arnaud JPE and Hofman, Albert and Krestin, Gabriel P and Niessen, Wiro J and Breteler, Monique MB and van der Lugt, Aad},
  journal={New England Journal of Medicine},
  volume={357},
  number={18},
  pages={1821--1828},
  year={2007},
  publisher={Mass Medical Soc}
}
@article{dice1945measures,
  title={Measures of the amount of ecologic association between species},
  author={Dice, Lee R},
  journal={Ecology},
  volume={26},
  number={3},
  pages={297--302},
  year={1945},
  publisher={JSTOR}
}
@inproceedings{davis2006relationship,
  title={The relationship between Precision-Recall and ROC curves},
  author={Davis, Jesse and Goadrich, Mark},
  booktitle={Proceedings of the 23rd international conference on Machine learning},
  pages={233--240},
  year={2006}
}
@misc{wang2023fremae,
      title={FreMAE: Fourier Transform Meets Masked Autoencoders for Medical Image Segmentation}, 
      author={Wenxuan Wang and Jing Wang and Chen Chen and Jianbo Jiao and Lichao Sun and Yuanxiu Cai and Shanshan Song and Jiangyun Li},
      year={2023},
      eprint={2304.10864},
      archivePrefix={arXiv},
      primaryClass={cs.CV}
}
@misc{gao2022convmae,
      title={ConvMAE: Masked Convolution Meets Masked Autoencoders}, 
      author={Peng Gao and Teli Ma and Hongsheng Li and Ziyi Lin and Jifeng Dai and Yu Qiao},
      year={2022},
      eprint={2205.03892},
      archivePrefix={arXiv},
      primaryClass={cs.CV}
}
@misc{he2021masked,
      title={Masked Autoencoders Are Scalable Vision Learners}, 
      author={Kaiming He and Xinlei Chen and Saining Xie and Yanghao Li and Piotr Dollár and Ross Girshick},
      year={2021},
      eprint={2111.06377},
      archivePrefix={arXiv},
      primaryClass={cs.CV}
}
@misc{ixii,   
   
    url = {https://brain-development.org/ixi-dataset/},   
    
       
}
@book{nussbaumer1981fast,
  title={The fast Fourier transform},
  author={Nussbaumer, Henri J and Nussbaumer, Henri J},
  year={1981},
  publisher={Springer}
}
@inproceedings{chen2019drop,
  title={Drop an octave: Reducing spatial redundancy in convolutional neural networks with octave convolution},
  author={Chen, Yunpeng and Fan, Haoqi and Xu, Bing and Yan, Zhicheng and Kalantidis, Yannis and Rohrbach, Marcus and Yan, Shuicheng and Feng, Jiashi},
  booktitle={Proceedings of the IEEE/CVF international conference on computer vision},
  pages={3435--3444},
  year={2019}
}
@article{kauffmann2014neural,
  title={The neural bases of spatial frequency processing during scene perception},
  author={Kauffmann, Louise and Ramano{\"e}l, Stephen and Peyrin, Carole},
  journal={Frontiers in integrative neuroscience},
  volume={8},
  pages={37},
  year={2014},
  publisher={Frontiers Media SA}
}
@article{10.1145/3464423,
author = {Fernando, Tharindu and Gammulle, Harshala and Denman, Simon and Sridharan, Sridha and Fookes, Clinton},
title = {Deep Learning for Medical Anomaly Detection – A Survey},
year = {2021},
issue_date = {September 2022},
publisher = {Association for Computing Machinery},
address = {New York, NY, USA},
volume = {54},
number = {7},
issn = {0360-0300},
url = {https://doi.org/10.1145/3464423},
doi = {10.1145/3464423},
abstract = {Machine learning–based medical anomaly detection is an important problem that has been extensively studied. Numerous approaches have been proposed across various medical application domains and we observe several similarities across these distinct applications. Despite this comparability, we observe a lack of structured organisation of these diverse research applications such that their advantages and limitations can be studied. The principal aim of this survey is to provide a thorough theoretical analysis of popular deep learning techniques in medical anomaly detection. In particular, we contribute a coherent and systematic review of state-of-the-art techniques, comparing and contrasting their architectural differences as well as training algorithms. Furthermore, we provide a comprehensive overview of deep model interpretation strategies that can be used to interpret model decisions. In addition, we outline the key limitations of existing deep medical anomaly detection techniques and propose key research directions for further investigation.},
journal = {ACM Comput. Surv.},
month = {jul},
articleno = {141},
numpages = {37},
keywords = {machine learning, anomaly detection, Deep learning, temporal analysis}
}
@article{bruno2015understanding,
  title={Understanding and confronting our mistakes: the epidemiology of error in radiology and strategies for error reduction},
  author={Bruno, Michael A and Walker, Eric A and Abujudeh, Hani H},
  journal={Radiographics},
  volume={35},
  number={6},
  pages={1668--1676},
  year={2015},
  publisher={Radiological Society of North America}
}

@article{Drew.2013,
 author = {Drew, Trafton and V{\~o}, Melissa L-H and Wolfe, Jeremy M.},
 year = {2013},
 title = {{The invisible gorilla strikes again: sustained inattentional blindness in expert observers}},
 pages = {1848--1853},
 volume = {24},
 number = {9},
 journal = {{Psychological science}},
 abstract = {Researchers have shown that people often miss the occurrence of an unexpected yet salient event if they are engaged in a different task, a phenomenon known as inattentional blindness. However, demonstrations of inattentional blindness have typically involved naive observers engaged in an unfamiliar task. What about expert searchers who have spent years honing their ability to detect small abnormalities in specific types of images? We asked 24 radiologists to perform a familiar lung-nodule detection task. A gorilla, 48 times the size of the average nodule, was inserted in the last case that was presented. Eighty-three percent of the radiologists did not see the gorilla. Eye tracking revealed that the majority of those who missed the gorilla looked directly at its location. Thus, even expert searchers, operating in their domain of expertise, are vulnerable to inattentional blindness.}
}

@article{Lundervold.2019,
 author = {Lundervold, Alexander Selvikv{\aa}g and Lundervold, Arvid},
 year = {2019},
 title = {{An overview of deep learning in medical imaging focusing on MRI}},
 pages = {102--127},
 volume = {29},
 number = {2},
 journal = {{Zeitschrift fur medizinische Physik}},
 abstract = {What has happened in machine learning lately, and what does it mean for the future of medical image analysis? Machine learning has witnessed a tremendous amount of attention over the last few years. The current boom started around 2009 when so-called deep artificial neural networks began outperforming other established models on a number of important benchmarks. Deep neural networks are now the state-of-the-art machine learning models across a variety of areas, from image analysis to natural language processing, and widely deployed in academia and industry. These developments have a huge potential for medical imaging technology, medical data analysis, medical diagnostics and healthcare in general, slowly being realized. We provide a short overview of recent advances and some associated challenges in machine learning applied to medical image processing and image analysis. As this has become a very broad and fast expanding field we will not survey the entire landscape of applications, but put particular focus on deep learning in MRI. Our aim is threefold: (i) give a brief introduction to deep learning with pointers to core references; (ii) indicate how deep learning has been applied to the entire MRI processing chain, from acquisition to image retrieval, from segmentation to disease prediction; (iii) provide a starting point for people interested in experimenting and perhaps contributing to the field of deep learning for medical imaging by pointing out good educational resources, state-of-the-art open-source code, and interesting sources of data and problems related medical imaging.}
}


 @article{Kim.2014,
author = {Kim, Young W. and Mansfield, Liem T.},
title = {Fool Me Twice: Delayed Diagnoses in Radiology With Emphasis on Perpetuated Errors},
journal = {American Journal of Roentgenology},
volume = {202},
number = {3},
pages = {465-470},
year = {2014},
doi = {10.2214/AJR.13.11493 },
    note ={PMID: 24555582 },

URL = { 
        https://doi.org/10.2214/AJR.13.11493
    
},
eprint = { 
        https://doi.org/10.2214/AJR.13.11493
    
}

}

@inproceedings{shortagerad,
author = {Mirza, Kashif and Vinayak, Sudhir and Khan, Tanveer and Shituma, Shikuku and Mugoma, Godfred},
year = {2012},
month = {11},
pages = {},
booktitle  = {Radiological Society of North America 2012 Scientific Assembly and Annual Meeting},
title = {Overcoming Shortage of Radiologists by Implementing a Cross-Border PACS, RIS and HIS in East Africa (EA) (Kenya, Uganda and Tanzania)}
}

@article{shen2017deep,
  title={Deep learning in medical image analysis},
  author={Shen, Dinggang and Wu, Guorong and Suk, Heung-Il},
  journal={Annual review of biomedical engineering},
  volume={19},
  pages={221},
  year={2017},
  publisher={NIH Public Access}
}

@article{ELLIS2022100068,
title = {Twelve key challenges in medical machine learning and solutions},
journal = {Intelligence-Based Medicine},
volume = {6},
pages = {100068},
year = {2022},
issn = {2666-5212},
author = {Randall J. Ellis and Ryan M. Sander and Alfonso Limon},
keywords = {Machine learning, Baseline models, Performance metrics, Imbalanced datasets, Model and label uncertainty, Reproducibility},
abstract = {The utility of machine learning in biomedicine is being investigated in various contexts, including for diagnostic and interpretive purposes for imaging modalities, quantifying disease risk, and processing text from physician and patient reports. To best facilitate the potential of machine learning, clinicians and computational scientists must inform one another about the nature of their clinical challenges and available methods for solving them, respectively. To this end, clinicians need to critically evaluate machine learning studies conducted to solve relevant problems in medicine. This article serves as a checklist for clinicians to understand and appraise machine learning studies and help facilitate productive conversations between the clinical and data science communities to improve human health.}
}

@article{karimi2020deep,
  title={Deep learning with noisy labels: Exploring techniques and remedies in medical image analysis},
  author={Karimi, Davood and Dou, Haoran and Warfield, Simon K and Gholipour, Ali},
  journal={Medical Image Analysis},
  volume={65},
  pages={101759},
  year={2020},
  publisher={Elsevier}
}

@article{Johnson.2019,
 author = {Johnson, Justin M. and Khoshgoftaar, Taghi M.},
 year = {2019},
 title = {{Survey on deep learning with class imbalance}},
 pages = {1--54},
 volume = {6},
 number = {1},
 journal = {{Journal of Big Data}},
 abstract = {The purpose of this study is to examine existing deep learning techniques for addressing class imbalanced data. Effective classification with imbalanced data is an important area of research, as high class imbalance is naturally inherent in many real-world applications, e.g., fraud detection and cancer detection. Moreover, highly imbalanced data poses added difficulty, as most learners will exhibit bias towards the majority class, and in extreme cases, may ignore the minority class altogether. Class imbalance has been studied thoroughly over the last two decades using traditional machine learning models, i.e. non-deep learning. Despite recent advances in deep learning, along with its increasing popularity, very little empirical work in the area of deep learning with class imbalance exists. Having achieved record-breaking performance results in several complex domains, investigating the use of deep neural networks for problems containing high levels of class imbalance is of great interest. Available studies regarding class imbalance and deep learning are surveyed in order to better understand the efficacy of deep learning when applied to class imbalanced data. This survey discusses the implementation details and experimental results for each study, and offers additional insight into their strengths and weaknesses. Several areas of focus include: data complexity, architectures tested, performance interpretation, ease of use, big data application, and generalization to other domains. We have found that research in this area is very limited, that most existing work focuses on computer vision tasks with convolutional neural networks, and that the effects of big data are rarely considered. Several traditional methods for class imbalance, e.g. data sampling and cost-sensitive learning, prove to be applicable in deep learning, while more advanced methods that exploit neural network feature learning abilities show promising results. The survey concludes with a discussion that highlights various gaps in deep learning from class imbalanced data for the purpose of guiding future research.}
}

@article{baur2021autoencoders,
  title={Autoencoders for unsupervised anomaly segmentation in brain MR images: a comparative study},
  author={Baur, Christoph and Denner, Stefan and Wiestler, Benedikt and Navab, Nassir and Albarqouni, Shadi},
  journal={Medical Image Analysis},
  pages={101952},
  year={2021},
  publisher={Elsevier}
}


@inproceedings{Chen.2018,
 author = {Chen, Xiaoran and Konukoglu, Ender},
 title = {{Unsupervised Detection of Lesions in Brain MRI using Constrained Adversarial Auto-encoders}},
 publisher = {PMLR},
 series = {{Proceedings of Machine Learning Research}},
 booktitle = {{International Conference on Medical Imaging with Deep Learning (MIDL)}},
 year = {2018}
}

@article{chen2020unsupervised,
  title={Unsupervised lesion detection via image restoration with a normative prior},
  author={Chen, Xiaoran and You, Suhang and Tezcan, Kerem Can and Konukoglu, Ender},
  journal={Medical image analysis},
  volume={64},
  pages={101713},
  year={2020},
  publisher={Elsevier}
}
@article{Schlegl.2019,
 author = {Schlegl, Thomas and Seeb{\"o}ck, Philipp and Waldstein, Sebastian M. and Langs, Georg and Schmidt-Erfurth, Ursula},
 year = {2019},
 title = {{f-AnoGAN: Fast unsupervised anomaly detection with generative adversarial networks}},
 pages = {30--44},
 volume = {54},
 journal = {{Medical image analysis}},
 abstract = {Obtaining expert labels in clinical imaging is difficult since exhaustive annotation is time-consuming. Furthermore, not all possibly relevant markers may be known and sufficiently well described a priori to even guide annotation. While supervised learning yields good results if expert labeled training data is available, the visual variability, and thus the vocabulary of findings, we can detect and exploit, is limited to the annotated lesions. Here, we present fast AnoGAN (f-AnoGAN), a generative adversarial network (GAN) based unsupervised learning approach capable of identifying anomalous images and image segments, that can serve as imaging biomarker candidates. We build a generative model of healthy training data, and propose and evaluate a fast mapping technique of new data to the GAN's latent space. The mapping is based on a trained encoder, and anomalies are detected via a combined anomaly score based on the building blocks of the trained model - comprising a discriminator feature residual error and an image reconstruction error. In the experiments on optical coherence tomography data, we compare the proposed method with alternative approaches, and provide comprehensive empirical evidence that f-AnoGAN outperforms alternative approaches and yields high anomaly detection accuracy. In addition, a visual Turing test with two retina experts showed that the generated images are indistinguishable from real normal retinal OCT images. The f-AnoGAN code is available at https://github.com/tSchlegl/f-AnoGAN.}
}


@article{wolleb2022diffusion,
  title={Diffusion Models for Medical Anomaly Detection},
  author={Wolleb, Julia and Bieder, Florentin and Sandk{\"u}hler, Robin and Cattin, Philippe C},
  journal={arXiv preprint arXiv:2203.04306   } ,
  year={2022}
}

@inproceedings{wyatt2022anoddpm,
  title={AnoDDPM: Anomaly Detection With Denoising Diffusion Probabilistic Models Using Simplex Noise},
  author={Wyatt, Julian and Leach, Adam and Schmon, Sebastian M and Willcocks, Chris G},
  booktitle={Proceedings of the IEEE/CVF Conference on Computer Vision and Pattern Recognition},
  pages={650--656},
  year={2022}
}

@article{pinaya2022fast,
  title={Fast Unsupervised Brain Anomaly Detection and Segmentation with Diffusion Models},
  author={Pinaya, Walter HL and Graham, Mark S and Gray, Robert and Da Costa, Pedro F and Tudosiu, Petru-Daniel and Wright, Paul and Mah, Yee H and MacKinnon, Andrew D and Teo, James T and Jager, Rolf and others},
  journal={arXiv preprint arXiv:2206.03461  } ,
  year={2022}
}

@inproceedings{sanchez2022healthy,
  title={What is healthy? generative counterfactual diffusion for lesion localization},
  author={Sanchez, Pedro and Kascenas, Antanas and Liu, Xiao and O’Neil, Alison Q and Tsaftaris, Sotirios A},
  booktitle={MICCAI Workshop on Deep Generative Models},
  pages={34--44},
  year={2022},
  organization={Springer}
}

@article{ho2020denoising,
  title={Denoising diffusion probabilistic models},
  author={Ho, Jonathan and Jain, Ajay and Abbeel, Pieter},
  journal={Advances in Neural Information Processing Systems},
  volume={33},
  pages={6840--6851},
  year={2020}
}

@article{dhariwal2021diffusion,
  title={Diffusion models beat gans on image synthesis},
  author={Dhariwal, Prafulla and Nichol, Alexander},
  journal={Advances in Neural Information Processing Systems},
  volume={34},
  pages={8780--8794},
  year={2021}
}
@inproceedings{rombach2022high,
  title={High-resolution image synthesis with latent diffusion models},
  author={Rombach, Robin and Blattmann, Andreas and Lorenz, Dominik and Esser, Patrick and Ommer, Bj{\"o}rn},
  booktitle={Proceedings of the IEEE/CVF Conference on Computer Vision and Pattern Recognition},
  pages={10684--10695},
  year={2022}
}

@inproceedings{kascenas2021denoising,
  title={Denoising Autoencoders for Unsupervised Anomaly Detection in Brain MRI},
  author={Kascenas, Antanas and Pugeault, Nicolas and O'Neil, Alison Q},
 publisher = {PMLR},
 series = {{Proceedings of Machine Learning Research}},
 booktitle = {{International Conference on Medical Imaging with Deep Learning (MIDL)}},
  year={2022}
}

@article{tschuchnig2022anomaly,
  title={Anomaly Detection in Medical Imaging-A Mini Review},
  author={Tschuchnig, Maximilian E and Gadermayr, Michael},
  journal={Data Science--Analytics and Applications},
  pages={33--38},
  year={2022},
  publisher={Springer}
}
@inproceedings{baur2018deep,
  title={Deep autoencoding models for unsupervised anomaly segmentation in brain MR images},
  author={Baur, Christoph and Wiestler, Benedikt and Albarqouni, Shadi and Navab, Nassir},
  booktitle={International MICCAI brainlesion workshop},
  pages={161--169},
  year={2018},
  organization={Springer}
}

@inproceedings{baur2020scale,
  title={Scale-space autoencoders for unsupervised anomaly segmentation in brain mri},
  author={Baur, Christoph and Wiestler, Benedikt and Albarqouni, Shadi and Navab, Nassir},
  booktitle={International Conference on Medical Image Computing and Computer-Assisted Intervention},
  pages={552--561},
  year={2020},
  organization={Springer}
}

@inproceedings{baur2020bayesian,
  title={Bayesian skip-autoencoders for unsupervised hyperintense anomaly detection in high resolution brain MRI},
  author={Baur, Christoph and Wiestler, Benedikt and Albarqouni, Shadi and Navab, Nassir},
  booktitle={2020 IEEE 17th International Symposium on Biomedical Imaging (ISBI)},
  pages={1905--1909},
  year={2020},
  organization={IEEE}
}


@inproceedings{zimmerer2018context,
  title={Context-encoding Variational Autoencoder for Unsupervised Anomaly Detection},
  author={Zimmerer, David and Kohl, Simon and Petersen, Jens and Isensee, Fabian and Maier-Hein, Klaus},
  booktitle={International Conference on Medical Imaging with Deep Learning--Extended Abstract Track},
  year={2019}
}
@article{bengs2021three,
  title={Three-dimensional deep learning with spatial erasing for unsupervised anomaly segmentation in brain MRI},
  author={Bengs, Marcel and Behrendt, Finn and Kr{\"u}ger, Julia and Opfer, Roland and Schlaefer, Alexander},
  journal={International journal of computer assisted radiology and surgery},
  volume={16},
  number={9},
  pages={1413--1423},
  year={2021},
  publisher={Springer}
}

@article{pinaya2022unsupervised,
  title={Unsupervised brain imaging 3D anomaly detection and segmentation with transformers},
  author={Pinaya, Walter HL and Tudosiu, Petru-Daniel and Gray, Robert and Rees, Geraint and Nachev, Parashkev and Ourselin, Sebastien and Cardoso, M Jorge},
  journal={Medical Image Analysis},
  volume={79},
  pages={102475},
  year={2022},
  publisher={Elsevier}
}

@inproceedings{behrendt2022capturing,
  title={Capturing Inter-Slice Dependencies of 3D Brain MRI-Scans for Unsupervised Anomaly Detection},
  author={Behrendt, Finn and Bengs, Marcel and Bhattacharya, Debayan and Kr{\"u}ger, Julia and Opfer, Roland and Schlaefer, Alexander},
  booktitle={Medical Imaging with Deep Learning},
  year={2022}
}

@article{SILVARODRIGUEZ2022102526,
title = {Constrained unsupervised anomaly segmentation},
journal = {Medical Image Analysis},
volume = {80},
pages = {102526},
year = {2022},
author = {Julio Silva-Rodríguez and Valery Naranjo and Jose Dolz},
keywords = {Unsupervised anomaly localization, Constraint segmentation, Brain lesions},
abstract = {Current unsupervised anomaly localization approaches rely on generative models to learn the distribution of normal images, which is later used to identify potential anomalous regions derived from errors on the reconstructed images. To address the limitations of residual-based anomaly localization, very recent literature has focused on attention maps, by integrating supervision on them in the form of homogenization constraints. In this work, we propose a novel formulation that addresses the problem in a more principled manner, leveraging well-known knowledge in constrained optimization. In particular, the equality constraint on the attention maps in prior work is replaced by an inequality constraint, which allows more flexibility. In addition, to address the limitations of penalty-based functions we employ an extension of the popular log-barrier methods to handle the constraint. Last, we propose an alternative regularization term that maximizes the Shannon entropy of the attention maps, reducing the amount of hyperparameters of the proposed model. Comprehensive experiments on two publicly available datasets on brain lesion segmentation demonstrate that the proposed approach substantially outperforms relevant literature, establishing new state-of-the-art results for unsupervised lesion segmentation.}
}

@INPROCEEDINGS{8852144,

  author={Sato, Kazuki and Hama, Kenta and Matsubara, Takashi and Uehara, Kuniaki},

  booktitle={2019 International Joint Conference on Neural Networks (IJCNN)}, 

  title={Predictable Uncertainty-Aware Unsupervised Deep Anomaly Segmentation}, 

  year={2019},

  volume={},

  number={},

  pages={1-7},

  doi={10.1109/IJCNN.2019.8852144  }}


@inproceedings{meissen2022challenging,
  title={Challenging current semi-supervised anomaly segmentation methods for brain MRI},
  author={Meissen, Felix and Kaissis, Georgios and Rueckert, Daniel},
  booktitle={International MICCAI brainlesion workshop},
  pages={63--74},
  year={2022},
  organization={Springer}
}

@article{elfwing2018sigmoid,
  title={Sigmoid-weighted linear units for neural network function approximation in reinforcement learning},
  author={Elfwing, Stefan and Uchibe, Eiji and Doya, Kenji},
  journal={Neural Networks},
  volume={107},
  pages={3--11},
  year={2018},
  publisher={Elsevier}
}

@inproceedings{perez2018film,
  title={Film: Visual reasoning with a general conditioning layer},
  author={Perez, Ethan and Strub, Florian and De Vries, Harm and Dumoulin, Vincent and Courville, Aaron},
  booktitle={Proceedings of the AAAI Conference on Artificial Intelligence},
  volume={32},
  year={2018}
}

@inproceedings{ronneberger2015u,
  title={U-net: Convolutional networks for biomedical image segmentation},
  author={Ronneberger, Olaf and Fischer, Philipp and Brox, Thomas},
  booktitle={International Conference on Medical image computing and computer-assisted intervention},
  pages={234--241},
  year={2015},
  organization={Springer}
}

@article{isensee2019automated,
  title={Automated brain extraction of multisequence MRI using artificial neural networks},
  author={Isensee, Fabian and Schell, Marianne and Pflueger, Irada and Brugnara, Gianluca and Bonekamp, David and Neuberger, Ulf and Wick, Antje and Schlemmer, Heinz-Peter and Heiland, Sabine and Wick, Wolfgang and others},
  journal={Human brain mapping},
  volume={40},
  number={17},
  pages={4952--4964},
  year={2019},
  publisher={Wiley Online Library}
}

@inproceedings{nguyen2021unsupervised,
  title={Unsupervised region-based anomaly detection in brain MRI with adversarial image inpainting},
  author={Nguyen, Bao and Feldman, Adam and Bethapudi, Sarath and Jennings, Andrew and Willcocks, Chris G},
  booktitle={2021 IEEE 18th international symposium on biomedical imaging (ISBI)},
  pages={1127--1131},
  year={2021},
  organization={IEEE}
}

@article{kawamoto2005improving,
  title={Improving clinical practice using clinical decision support systems: a systematic review of trials to identify features critical to success},
  author={Kawamoto, Kensaku and Houlihan, Caitlin A and Balas, E Andrew and Lobach, David F},
  journal={Bmj},
  volume={330},
  number={7494},
  pages={765},
  year={2005},
  publisher={British Medical Journal Publishing Group}
}

@article{lesjak2018novel,
  title={A novel public MR image dataset of multiple sclerosis patients with lesion segmentations based on multi-rater consensus},
  author={Lesjak, {\v{Z}}iga and Galimzianova, Alfiia and Koren, Ale{\v{s}} and Lukin, Matej and Pernu{\v{s}}, Franjo and Likar, Bo{\v{s}}tjan and {\v{S}}piclin, {\v{Z}}iga},
  journal={Neuroinformatics},
  volume={16},
  number={1},
  pages={51--63},
  year={2018},
  publisher={Springer}
}

@article{baid2021rsna,
  title={The rsna-asnr-miccai brats 2021 benchmark on brain tumor segmentation and radiogenomic classification},
  author={Baid, Ujjwal and Ghodasara, Satyam and Mohan, Suyash and Bilello, Michel and Calabrese, Evan and Colak, Errol and Farahani, Keyvan and Kalpathy-Cramer, Jayashree and Kitamura, Felipe C and Pati, Sarthak and others},
  journal={arXiv preprint arXiv:2107.02314  } ,
  year={2021}
}

@article{bakas2017advancing,
  title={Advancing the cancer genome atlas glioma MRI collections with expert segmentation labels and radiomic features},
  author={Bakas, Spyridon and Akbari, Hamed and Sotiras, Aristeidis and Bilello, Michel and Rozycki, Martin and Kirby, Justin S and Freymann, John B and Farahani, Keyvan and Davatzikos, Christos},
  journal={Scientific data},
  volume={4},
  number={1},
  pages={1--13},
  year={2017},
  publisher={Nature Publishing Group}
}

@article{menze2014multimodal,
  title={The multimodal brain tumor image segmentation benchmark (BRATS)},
  author={Menze, Bjoern H and Jakab, Andras and Bauer, Stefan and Kalpathy-Cramer, Jayashree and Farahani, Keyvan and Kirby, Justin and Burren, Yuliya and Porz, Nicole and Slotboom, Johannes and Wiest, Roland and others},
  journal={IEEE transactions on medical imaging},
  volume={34},
  number={10},
  pages={1993--2024},
  year={2014},
  publisher={IEEE}
}
@article{rohlfing2010sri24,
  title={The SRI24 multichannel atlas of normal adult human brain structure},
  author={Rohlfing, Torsten and Zahr, Natalie M and Sullivan, Edith V and Pfefferbaum, Adolf},
  journal={Human brain mapping},
  volume={31},
  number={5},
  pages={798--819},
  year={2010},
  publisher={Wiley Online Library}
}

@article{graham2022denoising,
  title={Denoising Diffusion Models for Out-of-Distribution Detection},
  author={Graham, Mark S and Pinaya, Walter HL and Tudosiu, Petru-Daniel and Nachev, Parashkev and Ourselin, Sebastien and Cardoso, M Jorge},
  journal={arXiv preprint arXiv:2211.07740},
  year={2022}
}

@article{raschkas_2018_mlxtend,
  author       = {Sebastian Raschka},
  title        = {MLxtend: Providing machine learning and data science 
                  utilities and extensions to Python’s  
                  scientific computing stack},
  journal      = {The Journal of Open Source Software},
  volume       = {3},
  number       = {24},
  month        = apr,
  year         = 2018,
  publisher    = {The Open Journal},
  doi          = {10.21105/joss.00638},
  url          = {http://joss.theoj.org/papers/10.21105/joss.00638}
}

@inproceedings{lugmayr2022repaint,
  title={Repaint: Inpainting using denoising diffusion probabilistic models},
  author={Lugmayr, Andreas and Danelljan, Martin and Romero, Andres and Yu, Fisher and Timofte, Radu and Van Gool, Luc},
  booktitle={Proceedings of the IEEE/CVF Conference on Computer Vision and Pattern Recognition},
  pages={11461--11471},
  year={2022}
}
@article{behrendt2023patched,
  title={Patched Diffusion Models for Unsupervised Anomaly Detection in Brain MRI},
  author={Behrendt, Finn and Bhattacharya, Debayan and Kr{\"u}ger, Julia and Opfer, Roland and Schlaefer, Alexander},
  journal={arXiv preprint arXiv:2303.03758},
  year={2023}
}
@article{ozdenizci2023restoring,
  title={Restoring vision in adverse weather conditions with patch-based denoising diffusion models},
  author={{\"O}zdenizci, Ozan and Legenstein, Robert},
  journal={IEEE Transactions on Pattern Analysis and Machine Intelligence},
  year={2023},
  publisher={IEEE}
}
\end{document}